\documentclass[twocolumn,showpacs,preprintnumbers]{revtex4}
\usepackage{graphicx}
\usepackage{dcolumn}
\usepackage{bm}
\input epsf

\begin{document}


\title{The role of curvature in the slowing down acceleration scenario}

\author{V\'ictor H. C\'ardenas}

\email{victor[at]dfa.uv.cl}

\author{Marco Rivera}

\email{marco.rivera[at]uv.cl}

\affiliation{Departamento de F\'{\i}sica y Astronom\'ia, Facultad de
Ciencias, Universidad de Valpara\'iso, Gran Breta\~na 1111,
Valpara\'iso, Chile}

\begin{abstract}
We introduce the curvature $\Omega_k$ as a new free parameter in the
Bayesian analysis using SNIa, BAO and CMB data, in a model with
variable equation of state parameter $w(z)$. We compare the results
using both the Constitution and Union 2 data sets, and also study
possible low redshift transitions in the deceleration parameter
$q(z)$. We found that, incorporating $\Omega_k$ in the analysis, it
is possible to make all the three observational probes consistent
using both SNIa data sets. Our results support dark energy evolution
at small redshift, and show that the tension between small and large
redshift probes is ameliorated. However, although the tension
decreases, it is still not possible to find a consensus set of
parameters that fit all the three data set using the
Chevalier-Polarski-Linder CPL parametrization.
\end{abstract}


\maketitle

\section{Introduction}

Since 1998 one of the biggest puzzles in cosmology is dark energy:
the unknown component responsible for the current accelerated
expansion of the universe \cite{dereview}. In its simpler form, this
can be described by a constant equation of state parameter $w=-1$,
corresponding to a cosmological constant, leading to the
$\Lambda$CDM model, the simplest model that fits a varied set of
observational data.

As the observations increase in precision, it is possible to reach
levels that enable us to discriminate between a cosmological
constant or a dynamical cosmological constant ($w=w(z)$). The source
of this dynamical dark energy could be both, a new field component
filling the universe, as a quintessence scalar field \cite{quinta},
or it can be produced by modifying gravity \cite{modgrav}.

In a recent paper \cite{Shafieloo:2009ti} the authors suggest that
current observational data could favor a scenario in which the
acceleration of the expansion has past a maximum value and is now
decelerating. The key point in deriving this conclusion is the use
of the Chevalier-Polarski-Linder (CPL) parametrization
\cite{Chevallier:2000qy}, \cite{Linder:2002et} for a dynamical
equation of state parameter
\begin{equation}\label{cpl}
 w(a)=w_0+(1-a)w_1,
\end{equation}
where $w_0$ and $w_1$ are constants to be fixed by observations and
$a$ is the scale factor. An updated analysis performed using recent
SNIa data was informed in \cite{Li:2010da} where similar conclusions
were derived. Both analysis assumed a flat universe.

Although it is often said that the inflationary scenario predicts a
flat universe \cite{inflation}, the situation has not been settled
down yet, in part because the observations are not conclusive and
because in theory $\Omega_k = 0$ is not the only solution
\cite{lin-mez}. In fact, the conclusion deduced from the analysis of
the observations using a $w=$constant, implying a model with
cosmological constant $\left|1+w\right|<0.06$, only works if we
assumed a flat universe. However, considering the curvature as a new
free parameter, it leads to an increment in the uncertainties in all
the others parameters. In particular, in the context of models with
dynamical dark energy, a number of works have discussed the
degeneracies between $w(z)$ and $\Omega_k$ \cite{linder05}
\cite{polran} \cite{HWS} \cite{clcoba} \cite{BFMV}.

It is the purpose of this paper to enhance the analysis of this
phenomenon, relaxing the assumption of a flat universe, considering
explicitly the geometric degeneracy between $\Omega_k$ and $w(z)$.
This enable us to explore new features considering the CPL
parametrization, enabling us to accommodate a similar behavior for
$q(z)$ using all the three observational test. This fact again shows
the sensitivity of the observational data to even smaller values of
the curvature parameter. The result found here shows that a single
trend for $q(z)$ is possible to draw using all the three
observational probes, ameliorating the tension between data sets.

The paper is organized as follows: in the next section we re-study
the flat universe case using both SNIa data sets
\cite{constitution}, \cite{Union2}. Then, we show how to generalize
the analysis considering a curved universe, in section III, which
are our main results. We end that section with a discussion.

\section{A flat Universe analysis: Union2 versus Constitution data}

The original analysis \cite{Shafieloo:2009ti} was performed using
the Consitution data set \cite{constitution} of SNIa. In this
section we reanalyze this data set along the BAO and CMB constraints
using the CPL parametrization. Then, we repeat the analysis using
the full Union2 data sample \cite{Union2}, looking for differences
in the consequences derived from these analysis. We end this section
discussing our results along the ones obtained in the recent work
\cite{Li:2010da}.

The luminosity distance in a flat universe is
\begin{equation}\label{dlzf}
d_L(z)=(1+z)\frac{c}{H_0}\int_0^z \frac{dz'}{E(z')},
\end{equation}
where
\begin{equation}\label{ezf}
E^2(z) = \Omega_m (1+z)^3 + (1-\Omega_m) e^{ 3 \int^z_0
\frac{1+w(z')}{1+z'} dz'}.
\end{equation}
The SNIa data give the distance modulus $\mu_{obs}(z)$ as a function
of redshift. We fit the SNIa with the cosmological model by
minimizing the $\chi^2$ value defined by
\begin{equation}
\chi_{SNIa}^2=\sum_{i=1}^{Np}\frac{[\mu(z_i)-\mu_{obs}(z_i)]^2}{\sigma_{\mu
i}^2},
\end{equation}
where  $\mu(z)\equiv 5\log_{10}[d_L(z)/Mpc]+25$ is the theoretical
value of the distance modulus, and $Np$ is the number of data
points.

The BAO data considered in our analysis is the distance ratio
obtained at $z=0.20$ and $z=0.35$ from the joint analysis of the 2dF
Galaxy Redsihft Survey and SDSS data \cite{bao2}, that can be
expressed as
\begin{equation}
\frac{D_V(0.35)}{D_V(0.20)}=1.736\pm 0.065,
\end{equation}
with
\begin{equation}
D_V(z_{BAO})=\bigg[\frac{z_{BAO}}{H(z_{BAO})}\bigg(\int_0^{z_{BAO}}\frac{dz}{H(z)}\bigg)^2\bigg]^{1/3}.
\end{equation}
We fit the cosmological model minimizing the $\chi^2$ defined by
\begin{equation}
\chi_{BAO}^2=\frac{[D_V(0.35)/D_V(0.20)-1.736]^2}{0.065^2}.
\end{equation}
A result from the combination of SNIa and BAO is given by a joint
analysis finding the best fit parameters that minimize
$\chi_{SNIa}^2+\chi_{BAO}^2$.

In addition, we incorporate to the analysis the CMB shift parameter
\cite{cmb2}, which is the reduce distance at $z_{ls}=1090$
\cite{Wang:2007mza}
\begin{equation}
R=\sqrt{\Omega_{m} H_0^2}r(z_{ls}) = 1.71\pm 0.019.
\end{equation}
Here we use the comoving distance $r(z)$ from the observer to
redshift $z$ which is related to the luminosity distance by
$d_L(z)=(1+z)r(z)$. In this case we compute $\chi^2$ by
\begin{equation}
\chi^2_{CMB}=\frac{[R-1.71]^2}{0.019^2}\;,
\end{equation}
to find out the result from CMB and the constraints from
SNIa+BAO+CMB are given by $\chi_{SNIa}^2+\chi_{BAO}^2+\chi_{CMB}^2$.

For the CPL parametrization, $w=w_0+w_1 z/(1+z)$, the second term in
the righ hand side of (\ref{ezf}) is
\begin{equation}
(1-\Omega_m)(1+z)^{3(1+w_0+w_1)}\exp
\left(-\frac{3w_1z}{1+z}\right).
\end{equation}

\begin{figure}[h!]
\centering \leavevmode\epsfysize=4.5cm \epsfbox{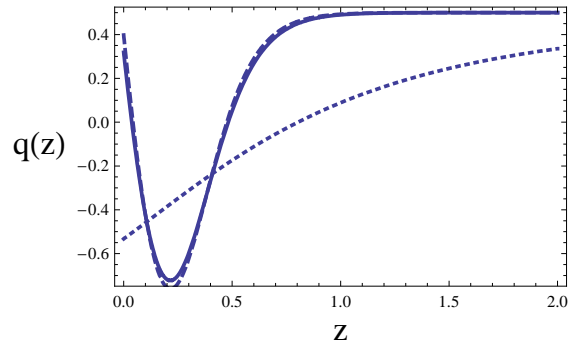}\\
\caption{\label{fig0} Using the Constitution data set
\cite{constitution} we plot the deceleration parameter reconstructed
 using the best fit values for three cases: only SNIa (continuous line),
 SNIa+BAO (dashed line) and SNIa+BAO+CMB (dotted line).}
 \end{figure}

\begin{figure}[h!]
\centering \leavevmode\epsfysize=7cm \epsfbox{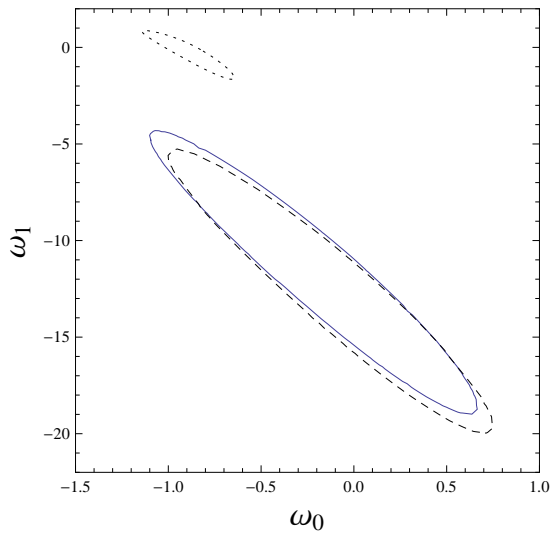}\\
\caption{\label{fig0b} Using the Constitution data set
\cite{constitution} we plot the $1 \sigma$ contours for the
$w_0-w_1$ parameters using the best fit values for three cases: only
SNIa (continuous line),
 SNIa+BAO (dashed line) and SNIa+BAO+CMB (dotted line).}
 \end{figure}

Using the Constitution sample consisting of $397$ data points,
assuming a flat universe, leads to the results shown in Table
\ref{tab:table0}. They are very similar to the ones informed in
\cite{Shafieloo:2009ti} and \cite{Wei:2009ry}.
\begin{table}[h!]
\caption{\label{tab:table0} The best fit values for the free
parameters using the Constitution data set in the case of a flat
universe model. See the related Fig. \ref{fig0}.}
\begin{ruledtabular}
\begin{tabular}{ccccc}
Data Set & $\chi^2_{min}$ & $\Omega_m$ & $w_0$ & $w_1$ \\
\hline
SN & 461.23 & 0.4519 & -0.2209 & -11.228 \\
SN+BAO & 461.60 & 0.4551 & -0.1253 & -12.255  \\
SN+BAO+CMB & 466.94 & 0.2578 & -0.9275 & -0.019 \\
\end{tabular}
\end{ruledtabular}
\end{table}
Given the number of degrees of freedom $N$, a good estimate of the
width of the $\chi^2$ distribution is $\sigma=\sqrt{2/N}$, which in
our case gives $\sigma \simeq 0.07$, a number taken into account to
discriminate statistical significance among different runs. Having
found these numbers, we can draw a profile of the deceleration
parameter $q=-\ddot{a}a/\dot{a}^2$ as a function of the redshift $z$
for each case. This is shown in Fig.\ref{fig0}. Clearly, the
incorporation of the CMB constraints into the fitting process, leads
to a completely different behavior for $q(z)$. The values of
$\chi^2_{min}$ show that the CPL parametrization is unable to fit
the data simultaneously at low and large redshift.
Because the best fit considering CMB gives a $\chi^2_{min}$ 5 points
higher than those considering SNIa and SNIa+BAO, indicates that CMB
data favor a different behavior at small redshift, as the one
observed in Fig.\ref{fig0}. Considering the reduced $\chi^2$ we
notice that $\chi^2_{red}=1.169$ for SNIa+BAO and
$\chi^2_{red}=1.182$ for SNIa+BAO+CMB, showing that using SNIa+BAO
gives a much better fit to the model than the one using CMB, and
again the inadequacy of the CPL parametrization of fitting data at
large and small redshift. The same can be concluded observing Figure
\ref{fig0b} where the confidence contours in the $w_0-w_1$ plane is
plotted . Notice that the ellipses corresponding to SNIa alone and
SNIa+BAO are almost identical, but the one considering CMB data does
not overlap the others, indicating the tension between these data
sets.

\begin{figure}[h!]
\centering \leavevmode\epsfysize=4.5cm \epsfbox{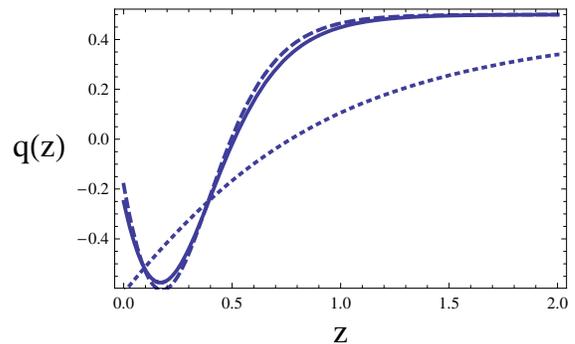}\\
\caption{\label{fig01} Using the Union 2 data set \cite{Union2} we
plot the deceleration parameter reconstructed using the best fit
values for three cases: only SNIa (continuous line),
 SNIa+BAO (dashed line) and SNIa+BAO+CMB (dotted line).}
 \end{figure}

Using the Union 2 set \cite{Union2}, the largest SNIa luminosity
distance sample currently available, consisting in $557$ type Ia
supernovae, the analysis leads to the results shown in Table
\ref{tab:table01}.
\begin{table}[h!]
\caption{\label{tab:table01} The best fit values for the free
parameters using the Union 2 data set in the case of a flat universe
model. See also Fig. \ref{fig01}.}
\begin{ruledtabular}
\begin{tabular}{ccccc}
Data Set & $\chi^2_{min}$ & $\Omega_m$ & $w_0$ & $w_1$ \\
\hline
SN & 541.43 & 0.4197 & -0.8632 & -5.490 \\
SN+BAO & 542.11 & 0.4281 & -0.7959 & -6.537  \\
SN+BAO+CMB & 543.91 & 0.2547 & -0.9979 & 0.190 \\
\end{tabular}
\end{ruledtabular}
\end{table}
In this case, $\sigma \simeq 0.06$, then again the numbers quoted in
the table have a trustable statistical meaning. However, by
comparing this with the results using the Constitution data set, we
observe that the difference between the case with and without the
CMB data is now smaller ($\simeq 1.8$), although still large
compared to $3\sigma = 0.18$. Plotting the deceleration parameter
using these numbers leads to Fig.\ref{fig01}.
Again, the case SNIa alone and SNIa+BAO are almost identical, but
the case including CMB data does not shown the rapid change at small
redshift as is also found using the Constitution data set. Similar
results were recently published in \cite{Li:2010da}. Meanwhile the
statistical analysis using the Constitution data set give
$\chi_{red}^2$ values greater than one, indicating bad fittings, the
ones obtained using the Union 2 sample, gives us over-fitting
values, $\chi_{red}^2<1$. This is important to notice, because an
over-fitting work indicates that probably we are fitting noise
instead of the actual relationship, and then more data would be
necessary to improve the predictive power of the model. Therefore,
it is necessary to perform a study that compares both data sets of
SNIa to shed some light on this point.

\begin{figure}[h!]
\centering \leavevmode\epsfysize=7cm \epsfbox{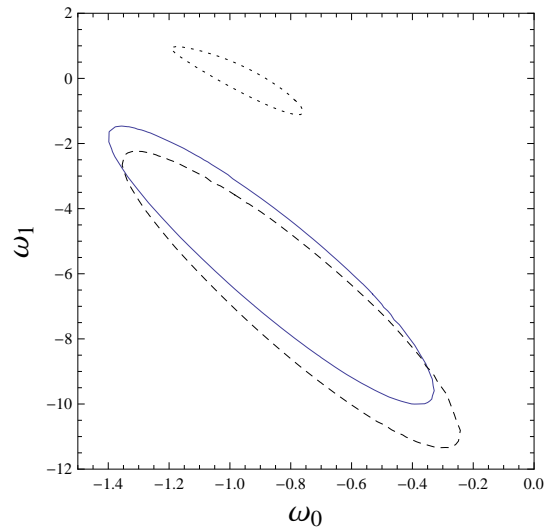}\\
\caption{\label{fig01b} Using the Union 2 data set \cite{Union2} we
plot the $1 \sigma$ contours for the $w_0-w_1$ parameters using the
best fit values for the three cases: only SNIa (continuous line),
 SNIa+BAO (dashed line) and SNIa+BAO+CMB (dotted line).}
 \end{figure}

This fact can also be viewed from Figure \ref{fig01b} where
confidence contours in the $w_0-w_1$ plane are plotted. Notice that
again, the ellipses corresponding to SNIa alone and SNIa+BAO are
similar, but the one considering SNIa+BAO+CMB points into a
completely different region. This confirm the previous claim that
the CPL parametrization is not adequate here.

\section{Relaxing the flat universe assumption}

The comoving distance from the observer to redshift $z$ is given by
\begin{eqnarray}\label{comdistance}
r(z)& = & \frac{c}{H_0}\frac{1}{\sqrt{-\Omega_k}}\sin(\sqrt{-\Omega_k}\Gamma(z)) \\
\Gamma(z)& = & \int_0^z \frac{dz'}{E(z')},\nonumber
\end{eqnarray}
where $\Omega_k=-k/H_0^2$, $k$ being the curvature constant and

\begin{eqnarray}\label{edez}
  E^2(z) & = & \Omega_m (1+z)^3+\Omega_k (1+z)^2 + \Omega_{de}f(z), \\
  f(z) & = & \exp \left\{ 3 \int^z_0 \frac{1+w(z')}{1+z'} dz'
  \right\},\nonumber
\end{eqnarray}
and $\Omega_{de}=1-\Omega_m - \Omega_k$.

For the CPL parametrization, $w=w_0+w_1 z/(1+z)$, the function
$f(z)$ in (\ref{edez}) leads to
\begin{equation}
f(z)=(1+z)^{3(1+w_0+w_1)}\exp \left(-\frac{3w_1z}{1+z}\right).
\end{equation}
Using the Constitution set we found the results displayed in table
\ref{tab:table1}. A direct comparison with the result of table
\ref{tab:table0}, where $\Omega_k=0$ was assumed, shows many
interesting changes. In general, all the values for $\chi^2_{min}$
are lower considering $\Omega_k$ a free parameter. This result is
expected, but the best fit values obtained are not. For example, in
the cases where SNIa and SNIa+BAO are considered, the best fit
values are completely different: $\Omega_k$ is around $-1$,
meanwhile the value for $w_1$ is four times less negative than in
the flat case. Also the matter density becomes almost twice the
value that in the flat case.
\begin{table}
\caption{\label{tab:table1} The best fit values for the free
parameters using the Constitution data set in the case of a non flat
universe model. See Fig. \ref{fig3}.}
\begin{ruledtabular}
\begin{tabular}{cccccc}
Data Set & $\chi^2_{min}$ & $\Omega_m$ & $w_0$ & $w_1$ & $\Omega_k$ \\
\hline
SN & 459.15 & 0.7049 & -0.4698 & -3.6307 & -0.9901 \\
SN+BAO & 459.37 & 0.6861 & -0.4868 & -2.8019 & -1.1534  \\
SN+BAO+CMB & 461.14 & 0.4934 & -0.1625 & -11.2574 & -0.0886 \\
\end{tabular}
\end{ruledtabular}
\end{table}

\begin{figure}[h]
\centering \leavevmode\epsfysize=4.5cm \epsfbox{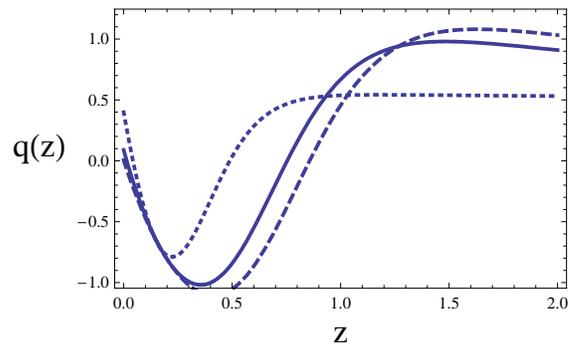}\\
\caption{\label{fig3} Using the Constitution data set
\cite{constitution} we plot the deceleration parameter reconstructed
using the best fit values for three cases: only SNIa (continuous
line), SNIa+BAO (dashed line) and SNIa+BAO+CMB (dotted line).}
\end{figure}

Using the best fit values of the parameters we plot $q(z)$ versus
$z$ for each case. We see from Fig.\ref{fig3}, that all the three
cases behaves in a similar way. This is one of the main result of
our work: the joint analysis SNIa+BAO+CMB using the curvature shows
qualitatively the same variation in $q(z)$ for small redshift than
those for SNIa alone or SNIa+BAO, a results that as far as we know
has not been informed previously.

The SNIa alone and SNIa+BAO look very similar to the already known,
with the exception that $q(z)$ is almost reaching its zero value
today $z=0$.
In each run, we have explored several different initial conditions,
to find the best fit values. Regarding the case where is used
SNIa+BAO+CMB, using a small value for $\Omega_k =-0.089$, enable us
to find a $q(z)$ whose form follows the same trend that the previous
cases. Here we must emphasize three important points: first, the
present value of the deceleration parameter is positive today
($z=0$), indicating that our universe is now decelerating, having
crossed to positive values at $z \simeq 0.1$. Second, the form of
$q(z)$ shows that even considering the CMB data, the incorporation
of the curvature to the analysis allows a transition in the
deceleration parameter at small redshift $z<0.3$. And finally,
although the SNIa+BAO+CMB case fit a $q(z)$ that change sign, the
profile of $q(z)$ reaches a minimum value later (at $z \simeq 0.2$)
than in the other two cases ($z \simeq 0.4$), crossing to negative
values later than the other cases and also in the SNIa+BAO+CMB case
the asymptotic value $q\simeq 1/2$ for $z>1$ is less than the one
found in the other cases $q\simeq 1$. However, this results does not
imply that the tension between low and high redshift data has
disappeared. In fact, by drawing confidence contours around the best
fit values in the $w_0-w_1$ plane shown in figure \ref{fig3b}, again
the ellipses at one sigma does not overlap at all, even those for
SNIa and SNIa+BAO.
\begin{figure}[h]
\centering \leavevmode\epsfysize=7cm \epsfbox{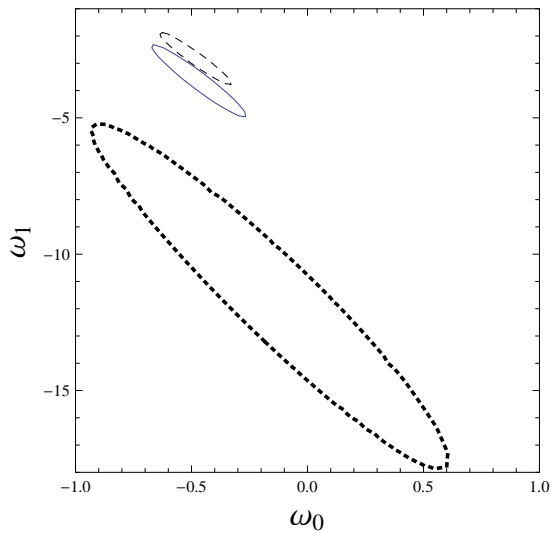}\\
\caption{\label{fig3b} Using the Constitution data set
\cite{constitution} we plot the $1 \sigma$ contours for the
$w_0-w_1$ parameters using the best fit values for three cases: only
SNIa (continuous line), SNIa+BAO (dashed line) and SNIa+BAO+CMB
(dotted line).}
\end{figure}

\begin{figure}[h!]
\centering \leavevmode\epsfysize=4.5cm \epsfbox{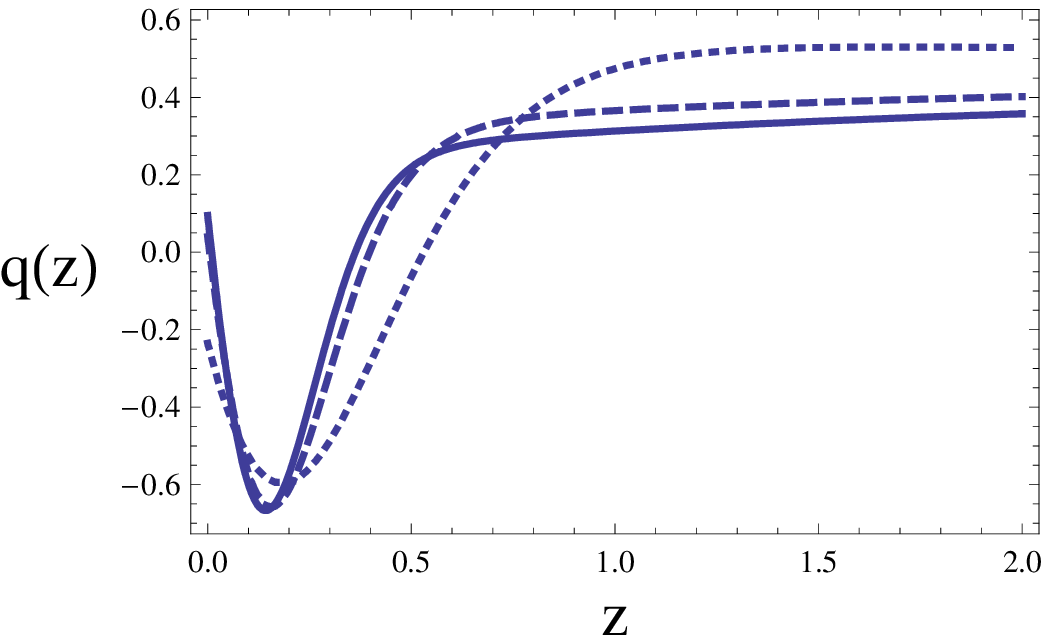}\\
\caption{\label{fig4} Using the Union 2 data set \cite{Union2} we
plot the deceleration parameter reconstructed using the best fit
values for three cases: only SNIa (continuos line), SNIa+BAO (dashed
line) and SNIa+BAO+CMB (dotted line).}
\end{figure}

Using the larger Union 2 data set \cite{Union2} we performed the
same analysis as before obtaining the best fit parameter shown in
table \ref{tab:table2}.
\begin{table}
\caption{\label{tab:table2} The best fit values for the free
parameters using the Union 2 data set in the case of a non flat
universe model. See Fig. \ref{fig4}.}
\begin{ruledtabular}
\begin{tabular}{cccccc}
Data Set & $\chi^2_{min}$ & $\Omega_m$ & $w_0$ & $w_1$ & $\Omega_k$ \\
\hline
SN & 541.05 & 0.2973 & -0.4362 & -19.343 & 0.3551 \\
SN+BAO & 541.91 & 0.3433 & -0.5488 & -14.531 & 0.2499 \\
SN+BAO+CMB & 542.18 & 0.4449 & -0.8180 & -5.1670 & -0.0744 \\

\end{tabular}
\end{ruledtabular}
\end{table}
The cases corresponding to SNIa data only and the joint analysis of
SNIa and BAO data, behaves similarly to the ones already displayed
using the Constitution data set as is evident in Fig.\ref{fig4}.
However in this case the change in sign of $q(z)$ take place later
than the ones discussed in Fig.\ref{fig3}.
Actually, this time all the three fits behave similarly, reaching a
minimum value at $z \simeq 0.2$. Again, the value of the
deceleration parameter today is almost zero for the SNIa alone and
SNIa+BAO cases, indicating that the acceleration of the expansion of
the universe had already ceased, and in the SNIa+BAO+CMB case $q$ is
small but still negative. However, as was also the case using the
Constitution set, the  concordance in the profile of $q(z)$ among
the observational probes is only apparent, because the one sigma
confidence contours in the plane $w_0-w_1$ indicate that although
SNIa and SNIa+BAO share a region in that phase space, the CMB data
spoiled the agreement.

Comparing table \ref{tab:table0} with \ref{tab:table1}, and
\ref{tab:table01} and \ref{tab:table2}, we observe that the
incorporation of the curvature parameter decreases the value of
$\chi^2_{min}$, indicating that a model with a non flat universe is
preferred, even with a small value. Also is clear from the analysis
that, both in the flat and non flat case, using the Constitution
data set, leads to a wrong fit $\chi^2_{red}>1$, meanwhile using the
Union 2 data set, leads to a over-fitting, $\chi^2_{red}<1$,
indicating the necessity of a further analysis of these SNIa data
sets.
\begin{figure}[h]
\centering \leavevmode\epsfysize=7cm \epsfbox{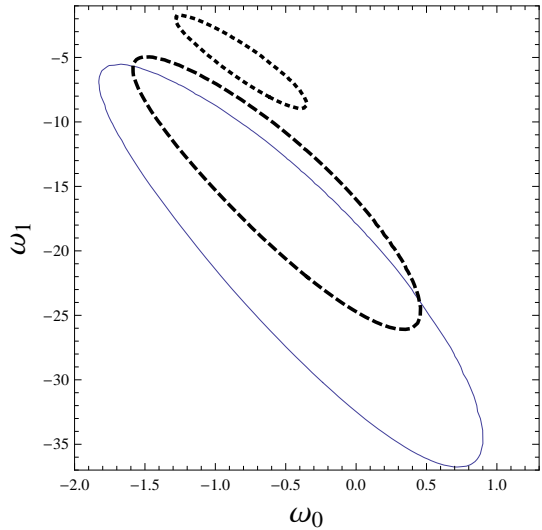}\\
\caption{\label{fig3b} Using the Union2 data set \cite{Union2} we
plot the $1 \sigma$ contours for the $w_0-w_1$ parameters using the
best fit values for three cases: only SNIa (continuous line),
SNIa+BAO (dashed line) and SNIa+BAO+CMB (dotted line).}
\end{figure}

One of the main result from this analysis is that, the incorporation
of the curvature as a new free parameter, enable us to accommodate a
similar behavior for $q(z)$ using all the three observational test.
This fact again shows the sensitivity of the observational data to
even smaller values (see Table \ref{tab:table2}) of the curvature
parameter. The result found here shows that a single trend for
$q(z)$ is possible to draw using all the three observational probes,
ameliorating the tension between data sets.


In this paper we have shown that incorporating the curvature
$\Omega_k$ as a free parameter in the Bayesian analysis using SNIa,
BAO and CMB, enable us to ameliorate the tension between low and
high redshift found in previous works
\cite{Shafieloo:2009ti},\cite{Li:2010da}. This is explicitly shown
using both the Constitution \cite{constitution} and the recent Union
2 \cite{Union2} data sets. Even considering small values for the
curvature parameter, it is possible to find a general behavior for
$q(z)$ at small redshift which shows a rapid variation in sign
between $z\simeq 0.5$ to $z\simeq 0$. Because the CPL ansatz
(\ref{cpl}) implicitly assumes that nothing special occurs for DE in
the redshift interval between $z=0$ to $z\simeq 2$, the trend found
in this work shows the incompatibility of the CPL parametrization in
describing the variation of $w(z)$ with redshift. This conclusion is
reinforced based on the lack of overlap among the confidence
contours in the three cases in the $w_0-w_1$ plane.

\section*{Acknowledgments}

The authors want to thanks Sergio del Campo for useful discussions.
This work was supported by Grant DIUV N$^{\circ}$ 13/09 from
Universidad de Valpara\'iso and FONDECYT Grant N$^{\circ}$ 1110230.



\begin{references}


\bibitem{dereview}
M. Sami, arXiv: 0904.3445 [hep-th]; J. Frieman, M. Turner and D.
Huterer, Ann. Rev. Astron. Astrophys. \textbf{46}, 385 (2008); T.
Padmanabhan, Gen. Rel. Grav. \textbf{40}, 529 (2008); M.S. Turner
and D. Huterer, J. Phys. Soc. Jap. \textbf{76}, 111015 (2007); N.
Straumann, Mod. Phys. Lett. \textbf{A21} 1083 (2006); E.J. Copeland,
M. Sami and S. Tsujikawa, Int. J. Mod. Phys. D \textbf{15}, 1753
(2006).

\bibitem{quinta}
C. Wetterich, Nucl. Phys. B \textbf{302}, 668 (1988); B. Ratra and
P.J.E. Peebles, Phys. Rev. D {\bf 37}, 3406 (1988); J.A. Frieman,
C.T. Hill, A. Stebbins and I. Waga, Phys. Rev. Lett. {\bf 75}, 2077
(1995) ; M.S. Turner and M. White, Phys. Rev. D {\bf 56}, R4439
(1997); R.R. Caldwell, R. Dave and P.J. Steinhardt, Phys. Rev. Lett.
{\bf 76}, 1582 (1998); A.R. Liddle and R.J. Scherrer, Phys. Rev. D
{\bf 59}, 023509 (1999); P.J. Steinhardt, L. Wang, and I. Zlatev,
Phys. Rev. D {\bf 59}, 123504 (1999).

\bibitem{modgrav}
G.R. Dvali, G. Gabadadze, and M. Porrati, Phys. Lett. B\textbf{485},
208 (2000); C. Deffayet, Phys. Lett. B\textbf{502}, 199 (2001); S.
Nojiri  and S. D. Odintsov, Phys. Rev. D 68, 123512 (2003); S.M.
Carroll, V. Duvvuri, M. Trodden, and M.S. Turner, Phys. Rev.
D\textbf{70}, 043528 (2004); S. Noriji and S.D. Odintsov,  Int. J.
Geom. Meth. Mod. Phys. \textbf{4}, 115 (2007); Y.S. Song, W. Hu, and
I. Sawicki, Phys. Rev. D\textbf{75}, 044004 (2007); S. Nojiri and S.
D. Odintsov, Phys. Rev. D 77, 026007 (2008).



\bibitem{Shafieloo:2009ti}
  A.~Shafieloo, V.~Sahni and A.~A.~Starobinsky,
  Phys.\ Rev.\  D {\bf 80}, 101301 (2009)
  [arXiv:0903.5141 [astro-ph.CO]].

\bibitem{Chevallier:2000qy}
  M.~Chevallier and D.~Polarski,
  Int.\ J.\ Mod.\ Phys.\  D {\bf 10}, 213 (2001)
  [arXiv:gr-qc/0009008].

\bibitem{Linder:2002et}
  E.~V.~Linder,
  Phys.\ Rev.\ Lett.\  {\bf 90}, 091301 (2003)
  [arXiv:astro-ph/0208512].

\bibitem{Li:2010da}
  Z.~Li, P.~Wu and H.~Yu,
  Phys.\ Lett.\  B {\bf 695}, 1 (2011)
  [arXiv:1011.1982 [gr-qc]].

\bibitem{Wei:2009ry}
  H.~Wei,
  Phys.\ Lett.\  B {\bf 687}, 286 (2010)
  [arXiv:0906.0828 [astro-ph.CO]].


\bibitem{inflation}
V. Mukhanov, {\it Physical Foundations of Cosmology}, Cambridge
(2005).

\bibitem{lin-mez}
A.Linde and A. Mezhlumian, Phys. Rev. D {\bf 52}, 6789 (1995); A.
Linde, D. Linde, A. Mezhlumian, Phys. Lett. B{\bf 345}, 203,(1995).

\bibitem{linder05}
E.V. Linder, Astropart. Phys. {\bf 24}, 391 (2005).

\bibitem{polran}
D. Polarski and A. Ranquet, Phys. Lett. {\bf B627}, 1 (2005).

\bibitem{HWS} Z
.~Y.~Huang, B.~Wang and R.~K.~Su, Int.\ J.\ Mod.\ Phys.\ A {\bf 22},
1819 (2007).

\bibitem{clcoba}
C. Clarkson, M. Cortes and B. Bassett, JCAP {\bf 0708}, 011 (2007).

\bibitem{BFMV}
G.~Barenboim, E.~Fernandez-Martinez, O.~Mena and L.~Verde, JCAP {\bf
1003}, 008 (2010).


\bibitem{constitution}
  M.~Hicken {\it et al.},
  Astrophys.\ J.\  {\bf 700}, 1097 (2009)
  [arXiv:0901.4804 [astro-ph.CO]].

\bibitem{Union2}
  R.~Amanullah {\it et al.},
  Astrophys.\ J.\  {\bf 716}, 712 (2010)
  [arXiv:1004.1711 [astro-ph.CO]].


\bibitem{bao2}
  B.~A.~Reid {\it et al.},
  Mon.\ Not.\ Roy.\ Astron.\ Soc.\  {\bf 404}, 60 (2010)
  [arXiv:0907.1659 [astro-ph.CO]];

  B.~A.~Reid {\it et al.}  [SDSS Collaboration],
  Mon.\ Not.\ Roy.\ Astron.\ Soc.\  {\bf 401}, 2148 (2010)
  [arXiv:0907.1660 [astro-ph.CO]].


\bibitem{cmb2}

  G.~Hinshaw {\it et al.}  [WMAP Collaboration],
  Astrophys.\ J.\ Suppl.\  {\bf 180}, 225 (2009)
  [arXiv:0803.0732 [astro-ph]];
  E.~Komatsu {\it et al.}  [WMAP Collaboration],
  Astrophys.\ J.\ Suppl.\  {\bf 180}, 330 (2009)
  [arXiv:0803.0547 [astro-ph]].

\bibitem{Wang:2007mza}
  Y.~Wang and P.~Mukherjee,
  Phys.\ Rev.\  D {\bf 76}, 103533 (2007)
  [arXiv:astro-ph/0703780].


\end{references}
\end{document}